\documentstyle[12pt]{article}
\input{tcilatex}
\begin{document}

\title{{\bf Mapping Integer Order Neumann Functions To Real Orders}}
\author{{\bf M.Mekhfi } \\
%EndAName
{\normalsize Laboratoire de Physique Mathematique,Es-senia 31100 Oran {\bf %
ALGERIE}}}
\date{}
\maketitle

\begin{abstract}
In a recent paper we unified Bessel functions of different orders .Here we
extend the unification to other linairely independant solutions to Bessel
equation ,Neumann's and Hankel's functions

{\footnotesize mekhfi@hotmail.com}
\end{abstract}

\newpage {}Neumann's functions $N_{p}(z)$ and Bessel functions $J_{p}(z)$
with p integer ,are the two linairely independant solutions to the Bessel
equation \cite{kn:Smirnov} \cite{kn:Watson} . 
\begin{equation}
z^{2}y^{^{\prime \prime }}(z)+zy^{^{\prime }}(z)+(z^{2}-p^{2})y(z)=0
\label{eq:1}
\end{equation}
\noindent When p is real it is known that $J_{p}(z)$ and $J_{-p}(z)$ are
also two linairely independant solutions of the above equation so
consequently we can rewrite $N_{p}(z)$ as a combination of

Bessel functions of positive and negative orders and the combination is as
follows 
\begin{equation}
N_{p}(z)=\frac{J_{p}(z)\,\,cosp\pi -J_{-p}(z)}{sinp\pi }  \label{eq:2}
\end{equation}
\noindent Bessel and Neumann's functions are the ''analog'' respectively of
the cosine and sine ,while Hankel functions $H_{p}^{(1)}$ and $H_{p}^{(2)}$
yet another linear combinations of $J_{p}(z)$ and $J_{-p}(z)$ are the
''analog'' of $e^{ipz}$ and $e^{-ipz}$. Hankel functions are related to $%
N_{p}(z)$ and $J_{p}(z)$ as. 
\begin{eqnarray}
J_{p}(z) &=&\frac{H_{p}^{(1)}(z)+H_{p}^{(2)}(z)}{2}  \nonumber \\
N_{p}(z) &=&\frac{H_{p}^{(1)}(z)-H_{p}^{(2)}(z)}{2i}  \label{eq:3}
\end{eqnarray}
\noindent where $H^{(1)}$ and $H^{(2)}$ are complex conjugate to each other.

In a recent paper \cite{kn:Wissale} we showed, for the first time , that
Bessel functions of real orders $J_{n+\lambda }$ $0\leq \lambda \leq 1$ and
Bessel functions $J_{n}$ $n\in Z$ of integer orders are no longer
disassociated objects , but one is the deformed\footnote{%
The deformation mechanism is described in \cite{Chamseddine}} version of the
other and as a consequence ,they are related to each other through the
unifying formula \footnote{%
We keep the notation of the partial derivative $\partial $ wherever and when
it acts on a single variable it is identified with a simple derivative $d.$} 
\begin{eqnarray}
\frac{J_{n+\lambda }(z)}{z^{n+\lambda }} &=&exp(-\lambda \sum_{z_{1}})\,\,%
\frac{J_{n}(z)}{z^{n}}  \label{eq:3'} \\
z^{n+\lambda }\,J_{n+\lambda }(z) &=&exp(\lambda
\sum_{z_{2}})\,\,z^{n}\,J_{n}(z)  \label{eq:3"} \\
\sum_{z_{1}} &=&\sum_{m\epsilon Z/0}\frac{\partial _{m}}{m}  \nonumber \\
\sum_{z_{2}} &=&\sum_{m\epsilon Z/0}(-1)^{m}\,\frac{\partial _{m}}{m} 
\nonumber \\
\partial _{|m|} &=&\frac{2\partial }{dz^{2}}\frac{2\partial }{dz^{2}}\cdots 
\frac{2\partial }{dz^{2}}.  \nonumber \\
\partial _{-|m|} &=&\int \frac{1}{2}dz^{2}.\,\int \frac{1}{2}%
dz^{2}..............\int \frac{1}{2}dz^{2}.  \nonumber
\end{eqnarray}
\noindent At this point it seems natural to extend this formula which
applies to $J_{n},$ to the second linairely independant solution $N_{n}$.The
apparently straigthforward way to do would be to use formula ~ \ref{eq:3'} 
\ref{eq:3"} and then apply the mapping operator $e^{\pm \lambda \sum }$ ,
but unfortunately our formula applies to the reduced Bessel function $%
J_{n}(z)/z^{n}$ rather than directly to $J_{n}(z)$ and this make our formula
~of no practical use for our purpose.In the following we will propose a nice
way to relate $N_{n+\lambda }$ to $N_{n}$ .

Let us recall first that to Bessel functions one can associate polynomials $%
O_{n}(z).$ These are Neumann polynomials .The relationship between them is
performed through the known formula.

\begin{eqnarray}
\frac{1}{t-z} &=&\sum_{-\infty }^{+\infty }J_{n}(z)\,\,A_{n}(t)  \label{eq:4}
\\
with\,\,\,\,\,O_{n}(z) &=&A_{n}+(-1)^{n}A_{-n}  \nonumber
\end{eqnarray}

The first step towards the solution of our problem is to show that if , to
Bessel functions one can associate Neumann polynomials $O_{n}(z),$ one can
equally well associate to reduced Bessel functions something and these
associated objects are precisely Neumann function $N_{n}$ $(z)$ . \noindent
We are not sure if a formula which relates reduced Bessel functions to
Neumann functions ,similar to the above one \ref{eq:4}, already exists in
the literature ,so we prefer to work it out first.In doing this we largely
follow a method which dates back to Sonine \cite{kn:Watson} and adapts it to
the present need.Let us first state the result.Let $\Omega (\tau )$ be an an
arbitrary function of $\tau $ and if $\Omega (\tau )=x$ ,let $\tau =\mho
(x)\,\,$so that $\mho \,\,$ is the inverse function to $\Omega $ .Denote the
function and the function we associate to it respectively $Z_{n}(z)\,\,$ and 
$A_{n}(t)\,\,$ and define them through their integral representations as
follows 
\begin{eqnarray*}
Z_{n}(z) &=&\frac{1}{2\pi i}\oint e^{-z^{2}\Omega +\frac{\tau }{2}}\frac{%
d\tau }{\tau ^{n+1}} \\
A_{n}(t) &=&\int_{0}^{\infty }e^{t^{2}x-\frac{\mho (x)}{2}}\,\,(\mho
(x))^{n}dx
\end{eqnarray*}
\noindent then it is not that difficult to prove the existence of a formula
relating both functions 
\begin{eqnarray}
\frac{1}{t^{2}-z^{2}} &=&-\sum_{-\infty }^{+\infty }Z_{n}(z)\,\,A_{n}(t)
\label{eq:5} \\
Re\,\,z^{2}\,\, &>&\,\,Re\,\,t^{2}  \nonumber
\end{eqnarray}
\noindent \noindent To prove the result \ref{eq:5} , suppose for any given
positive value of x that $\mid \tau \mid >\mid \mho (x)\mid $ on a closed
curve ${\cal C}$ surrounding the origin and the point z and that $\mid \tau
\mid <\mid \mho (x)\mid $ on a closed ${\it c}$ surrounding the origin but
not enclosing the point z .Then compute the series 
\begin{eqnarray*}
\sum_{n=-\infty }^{+\infty }Z_{n}A_{n} &=&\frac{1}{2\pi i}%
\sum_{n=0}^{+\infty }\oint_{{\cal C}}\int_{0}^{\infty }exp(-z^{2}\Omega
(\tau )+t^{2}x+\frac{1}{2}(\tau -\mho (x))\,\,\frac{\mho (x)^{n}}{\tau ^{n+1}%
}dx\,\,d\tau \\
&+&\frac{1}{2\pi i}\sum_{n=0}^{+\infty }\oint_{c}\int_{0}^{\infty
}exp(-z^{2}\Omega (\tau )+t^{2}x+\frac{1}{2}(\tau -\mho (x))\,\,\frac{\tau
^{n}}{\mho (x)^{n+1}}dx\,\,d\tau \\
&=&\frac{1}{2\pi i}\int_{0}^{\infty }(\oint_{{\cal C}}-\oint_{c})\frac{%
exp(-z^{2}\Omega (\tau )+t^{2}x+\frac{1}{2}(\tau -\mho (x)}{\tau -\mho (x)}%
d\tau \,dx. \\
&=&\frac{1}{2\pi i}\int_{0}^{\infty }\oint_{z}\frac{exp(-z^{2}\Omega (\tau
)+t^{2}x+\frac{1}{2}(\tau -\mho (x)}{\tau -\mho (x)}d\tau \,dx. \\
&=&\int_{0}^{\infty }e^{(t^{2}-z^{2})x}\,dx \\
&=&-\frac{1}{t^{2}-z^{2}}
\end{eqnarray*}
\noindent In going to the third line,we perform the summations $\frac{1}{%
\tau }\sum_{n=0}^{+\infty }\left( \frac{\mho (x)}{\tau }\right) ^{n}=\frac{1%
}{\tau -\mho (x)}$ ($\tau $ on the path ${\cal C}$ )and $\frac{1}{\mho }%
\sum_{n=0}^{+\infty }\left( \frac{\tau }{\mho (x)}\right) ^{n}=-\frac{1}{%
\tau -\mho (x)}$ ( $\tau $ on the path ${\it c}$ ) which are convergent
according to the conditions above .For the case of interest we have $%
Z_{n}(z)=\frac{J_{n}(z)}{z^{n}}\,\,$ , $\Omega (\tau )=\frac{1}{2\tau }$ and 
$\mho (x)=\frac{1}{2x}\,\,$ .In this case the function associated to the
reduced Bessel function has the integral form . 
\begin{equation}
A_{n}(t)=\frac{1}{2}\,\int_{0}^{\infty }e^{t^{2}\frac{x}{2}-\frac{1}{2x}}\,\,%
\frac{dx}{x^{n}}  \label{eq:7}
\end{equation}

The integral defining the function $\,A_{n}(t)\,$ is easily identified \cite
{kn:grad} to the $K_{n}(t)$ function which is the Hankel function of
imaginary argument $K_{\nu }(t)=\frac{\pi i}{2}e^{i\frac{\pi }{2}\nu
i}H_{-\nu }^{(1)}(it)$. 
\begin{eqnarray*}
\frac{1}{2}\int_{0}^{\infty }e^{-\frac{1}{2x}+\frac{t^{2}x}{2}}\frac{dx}{%
x^{n}} &=&\frac{K_{-n+1}(it)}{(it)^{-n+1}}=\frac{K_{-n+1}(-it)}{(-it)^{-n+1}}
\\
with &\mid &arg\,\,t\mid <\pi /2\,\,and\,\,Re\,\,t^{2}>0
\end{eqnarray*}

Equation \ref{eq:5} then \noindent relates reduced Bessel funnctions to
Hankel functions which are associated to them 
\begin{equation}
\frac{2}{\pi }\frac{i}{t^{2}-z^{2}}=\sum_{-\infty }^{+\infty }\frac{J_{n}(z)%
}{z^{n}}\,\,t^{n-1}H_{n-1}^{(1)}(t).  \label{eq:9}
\end{equation}
\noindent In writing this formula we have used the property of the Hankel
functions that $H_{-n}=(-1)^{n}H_{n}$ .

At this point if we restrict ourselves to t and z reals and take the complex
conjugate of the above formula we get another formula 
\begin{equation}
\frac{2}{\pi }\frac{i}{t^{2}-z^{2}}=-\sum_{-\infty }^{+\infty }\frac{J_{n}(z)%
}{z^{n}}\,\,t^{n-1}H_{n-1}^{(2)}(t).  \label{eq:10}
\end{equation}
\noindent where, this time, we use the property $H_{\nu
}^{(1)}(z)^{*}\,=\,H_{\nu }^{(2)}(z^{*})$ . Adding \ref{eq:9} to \ref{eq:10}
and using the definition of Neumann function in term of Hankel functions \ref
{eq:3} we get the desired result 
\begin{equation}
\frac{2}{\pi }\frac{1}{t^{2}-z^{2}}=\sum_{-\infty }^{+\infty }\frac{J_{n}(z)%
}{z^{n}}\,\,t^{n-1}N_{n-1}(t).  \label{eq:11}
\end{equation}
\noindent The particular form of the left hand side of this equation (a
happy event) and the presence of reduced Bessel functions on the right hand
side of the equation allow a straigthforward and quite elegant application
to Neumann 's function of the unifying formula \ref{eq:3'} ( valid for
reduced Bessel functions ) .In fact we can perform a succession of valid
operations . 
\begin{eqnarray}
\frac{2}{\pi }exp(-\lambda \sum_{z_{1}})\frac{1}{t^{2}-z^{2}}
&=&\sum_{-\infty }^{+\infty }exp(-\lambda \sum_{z_{1}})\frac{J_{n}(z)}{z^{n}}%
\,\,t^{n-1}N_{n-1}(t)  \label{eq:12} \\
\frac{2}{\pi }exp(-\lambda \sum_{t})\frac{1}{t^{2}-z^{2}} &=&\sum_{-\infty
}^{+\infty }\frac{J_{n+\lambda }(z)}{z^{n+\lambda }}\,\,t^{n-1}N_{n-1}(t) 
\nonumber \\
\frac{2}{\pi }\frac{1}{t^{2}-z^{2}} &=&\sum_{-\infty }^{+\infty }\frac{%
J_{n+\lambda }(z)}{z^{n+\lambda }}\,\,exp(\lambda \sum_{t})t^{n-1}N_{n-1}(t)
\nonumber \\
\frac{2}{\pi }\frac{1}{t^{2}-z^{2}} &=&\sum_{-\infty }^{+\infty }\frac{%
J_{n+\lambda }(z)}{z^{n+\lambda }}\,\,t^{n+\lambda -1}N_{n+\lambda -1}(t) 
\nonumber
\end{eqnarray}
\noindent Where the capital sigma operator with the t variable acquires an
extra factor $(-1)^{m}$ . 
\begin{equation}
\sum_{t}=\sum_{m\epsilon Z/0}(-1)^{m}\frac{\partial _{m}}{m}  \label{eq:13}
\end{equation}
\noindent This extra factor comes from the identity. 
\begin{equation}
\frac{\partial }{\partial z^{2}}\,\,\frac{1}{t^{2}-z^{2}}=-\frac{\partial }{%
\partial t^{2}}\,\,\frac{1}{t^{2}-z^{2}}  \label{eq:14}
\end{equation}
\noindent Note that in the last line of \ref{eq:12} we identified the result
of the action of the operator $exp(\lambda \sum_{t})$ on Neumann functions
of integer orders with Neumann function of orders $n+\lambda $ as for
reduced Bessel functions.That is we put 
\begin{equation}
t^{n+\lambda }N_{n+\lambda }(t)=exp(\lambda \sum_{t})\,\,t^{n}N_{n}(t)
\label{eq:15}
\end{equation}

It remains to show \ref{eq:15}, that is ,the result of the mapping operation
by the operator $e^{\lambda }\sum $ is indeed Neumann functions of real
orders $n+\lambda $ and not something else.To see this let us proceed as
follows 
\begin{eqnarray}
\frac{2}{\pi }\frac{1}{t^{2}-z^{2}} &=&\sum_{-\infty }^{+\infty }\frac{%
J_{n+\lambda }(z)}{z^{n+\lambda }}\,\,A_{n,\lambda }=\sum_{-\infty
}^{+\infty }\frac{J_{n+\lambda -m}(z)}{z^{n+\lambda -m}}\,\,A_{n,\lambda -m}
\label{eq:16} \\
&=&\sum_{-\infty }^{+\infty }\frac{J_{n+\lambda }(z)}{z^{n+\lambda }}%
\,\,A_{n+m,\lambda -m}  \nonumber
\end{eqnarray}
\noindent In the first line we use the fact that the sum is independant of $%
\lambda $ and hence change the parameter $\lambda \,\,\rightarrow \lambda -m$
.In the second line we absorb the index m as we sum over all $m\,\epsilon
\,Z\,$ .From \ref{eq:16} we infer the symmetry property of the $A_{n,\lambda
}$%
\begin{equation}
A_{n,\lambda }=A_{n+m,\lambda -m}\forall \,\,\lambda ,m  \label{eq:17}
\end{equation}
\noindent This symmetry means that $A_{n,\lambda }$ only depends on the
combination $n+\lambda $ that is $A_{n,\lambda }=A_{n+\lambda }$ .This
remark together with the fact that $A_{n+\lambda }\mid _{\lambda =0}=$ $%
\,\,t^{n}N_{n}(t)$ leads to the identification $\ref{eq:15}$as we
anticipated in the last line of $\ref{eq:12}$

\noindent Formula $\ref{eq:15}$ is the main result of the paper .It shows
that Neumann functions are unified irrespective of their orders being
integer or real .They are unified through the same formula as reduced Bessel
functions $z^{n}\,J_{n}(z)$ in \ref{eq:3"}.

Note that we could have already applied the procedure to Hankel function's
first by starting from the identities ~\ref{eq:9} and .\ref{eq:10}. In this
case we could have obtained a similar unifying formulas ( i=1,2 ). 
\begin{equation}
t^{n+\lambda }H_{n+\lambda }^{(i)}(t)=exp(\lambda
\sum_{t})\,\,t^{n}H_{n}^{(i)}(t)  \label{eq:18}
\end{equation}
\noindent Let us note that our formula $\ref{eq:18}$ for the evolution of
Hankel functions could be used to infer the evolution equation for reduced
Bessel function $z^{n}J_{n}$ by using the defining equation $\ref{eq:3}$ of
J'$^{s}$ in terms of H'$^{s}$ . By simple inspection one can see that the
result we get is indeed equation $\ref{eq:3"}$.

\end{document}